\documentclass[aps,prd,preprint,groupedaddress]{revtex4-1}
\usepackage[utf8]{inputenc}
\usepackage{amsmath, amsfonts, mathtext, amssymb, mathrsfs,subfigure,wrapfig}
\usepackage[utf8]{inputenc}
\usepackage{graphicx} 
\usepackage{color}
\usepackage{textcomp}
\input epsf

\newcommand*{\No}{\textnumero}

\newcommand{\fr}[2]{\frac{#1}{#2}}
\newcommand{\bra}[1]{\left( #1 \right)}  
\newcommand{\brb}[1]{\left[ #1 \right]}  
  
\newcommand{\mF}{\mathcal{F}}
\newcommand{\mG}{\mathcal{G}}
\newcommand{\mH}{\mathcal{H}}
\newcommand{\mL}{\mathcal{L}}
\newcommand{\Ga}{\Gamma}
\newcommand{\na}{\nabla}

\begin{document}

\title{Stable black holes in shift-symmetric Horndeski theories}

\author{Daria A. Tretyakova}
\email[]{daria.tretiakova@urfu.ru}
\affiliation{Institute for Natural Sciences, Ural Federal University, Lenin av. 51, Yekaterinburg, 620083, Russia}

\author{Kazufumi Takahashi}
\email[]{ktakahashi@resceu.s.u-tokyo.ac.jp}
\affiliation{Research Center for the Early Universe (RESCEU), Graduate School of Science, The University of Tokyo, Tokyo 113-0033, Japan}
\affiliation{Department of Physics, Graduate School of Science, The University of Tokyo, Tokyo 113-0033, Japan}

\date{\today}

\begin{abstract}
In shift-symmetric Horndeski theories, a static and spherically symmetric black hole can support linearly time-dependent scalar hair.
However, it was shown that such a solution generically suffers from ghost or gradient instability in the vicinity of the horizon.
In the present paper, we explore the possibility to avoid the instability, and present a new example of theory and its black hole solution with a linearly time-dependent scalar configuration.
We also discuss the stability of solutions with static scalar hair for a special case where nonminimal derivative coupling to the Einstein tensor appears.
\end{abstract}

\maketitle

\section{Introduction}

Scalar-tensor gravity is a widely accepted alternative to general relativity (GR). The most general scalar-tensor theory that yields second-order Euler-Lagrange equations was proposed by Horndeski~\cite{Horndeski:1974wa}.
This second-order nature of field equations is desirable since it trivially circumvents so-called Ostrogradsky ghosts associated with higher-order derivatives~\cite{Woodard:2015zca}.
The same result was also obtained by studying Galileons~\cite{Deffayet:2009wt, Deffayet:2009mn,Deffayet:2011gz,Kobayashi:2011nu}. The Galileon model is a ghost-free scalar effective field theory containing higher derivative terms in the action.
The resulting framework encompasses GR and many other modified theories of gravity, such as the Brans-Dicke theory and $f(R)$ theories.

In this paper, we consider the shift-symmetric subclass of the Horndeski Lagrangian, consisting of four parts~$\mathcal{L}_i~(i = 2, 3, 4, 5)$. Each part is characterized by an arbitrary function~$G_i(X)$ that depends on the canonical kinetic term~$X$ of the scalar field~$\phi$.
These functions manifest themselves nontrivially in cosmological solutions~\cite{Martin-Moruno:2015bda,Starobinsky:2016kua}, wormhole configurations~\cite{Korolev:2014hwa} and black hole (BH) solutions~\cite{Rinaldi:2012vy,Babichev:2013cya,Anabalon:2013oea,Minamitsuji:2013ura,Kobayashi:2014eva,Charmousis:2015aya} (see Ref.~\cite{Babichev:2016rlq} for a review).
An intriguing feature of the shift-symmetric Horndeski class is that it allows static BH solutions with a linearly time-dependent scalar configuration~\cite{Babichev:2013cya}.
This happens because $\phi$ appears only with derivatives in the field equations in the shift-symmetric Horndeski theories.
An especially interesting BH solution is the Schwarzschild-de Sitter (dS) metric with a nontrivial scalar profile, in which the vacuum cosmological constant is totally screened. Such a solution is called stealth BH since it cannot be distinguished from the one in GR at least at the background level. 
It should be noted that most of the solutions found so far are obtained within an even narrower subclass of the shift-symmetric Horndeski theories, i.e., a class of theories having reflection symmetry of the scalar field\footnote{BH solutions in a shift-symmetric Horndeski theory with $G_3\propto X$ have been found in Ref.~\cite{Babichev:2016fbg}.}.
Under the requirement of the reflection symmetry, the two terms~$\mathcal{L}_3$ and $\mathcal{L}_5$ that contain odd numbers of $\phi$ must vanish.
This subclass includes the theory with nonminimal derivative coupling to the Einstein tensor~[see Eqs.~\eqref{dc2Gmn}, \eqref{ac}].
Throughout the present paper, we focus on such shift- and reflection-symmetric Horndeski theories.

The stability of BH solutions in the Horndeski theory has been investigated in recent works.
For static and spherically symmetric metric and scalar configurations in generic Horndeski theories, the linear odd- and even-parity perturbations were studied in Refs.~\cite{Kobayashi:2012kh} and \cite{Kobayashi:2014wsa}, respectively. For solutions with a linearly time-dependent scalar field in the shift- and reflection-symmetric Horndeski theories, the odd-parity perturbation analysis was performed in Ref.~\cite{Ogawa:2015pea}. It was later extended to the case of generic shift-symmetric Horndeski theories in Ref.~\cite{Takahashi:2016dnv}.
The authors of Refs.~\cite{Ogawa:2015pea,Takahashi:2016dnv} stated that a BH solution with a time-dependent scalar profile generically suffers from ghost or gradient instability in the vicinity of the horizon, while there are some loopholes~\cite{Takahashi:2016dnv}.
The similar instability appears in higher dimensions, as shown in Ref.~\cite{Takahashi:2015pad} in the framework of the so-called Lovelock-Galileon theory.
Importantly, the generic instability does not arise if the scalar field is static, though this does not necessarily mean the solution is stable.

In the present paper, we investigate the possibility of avoiding the generic instability.
We show a new example of potentially stable BH with a linearly time-dependent scalar profile.
For a BH with static scalar hair, we focus on a theory with nonminimal derivative coupling to the Einstein tensor and obtain the parameter region allowed by the stability requirements.

This paper is organized as follows.
In the next section, we present the shift- and reflection-symmetric Horndeski theories and review hairy BH solutions within this framework.
In Sec.~\ref{bhs}, we discuss the linear stability of the BH solutions.
The cases of time-dependent and static scalar field are treated separately.
Finally, we conclude in Sec.~\ref{conc}.

\section{Shift- and reflection-symmetric Horndeski theories and background solutions} \label{background}

Let us start from the following action within the Horndeski framework:
\begin{equation}
S=\int d^4x\sqrt{-g}(\mathcal{L}_2+\mathcal{L}_4), \label{lag}
\end{equation}
where
\begin{eqnarray}
\mathcal{L}_2&=& G_2(X), \\
\mathcal{L}_4&=& G_4(X)R +G_{4X}[(\Box\phi)^2-(\nabla_{\mu}\nabla_{\nu}\phi)^2].
\end{eqnarray}
Here $R$ is the scalar curvature, $X=-(\nabla_\mu \phi)^2/2$ is the canonical kinetic term of the scalar field, $G_2$ and $G_4$ are arbitrary functions of $X$, and $G_{4X}\equiv d G_4/d X$.
This theory is invariant under the shift~$\phi \to \phi + c $ (with $c$ being an arbitrary real constant) and the reflection~$\phi\to-\phi$.
Within this framework, we study static and spherically symmetric BH solutions  with the ansatz adopted in Refs.~\cite{Babichev:2013cya,Kobayashi:2014eva,Charmousis:2015aya} having the nontrivial scalar profile of the form
\begin{eqnarray}
&&ds^2=-h(r)dt^2+\frac{dr^2}{f(r)}+r^2(d\theta^2+\sin^2 \theta d\varphi^2), \label{ans} \\
&&\phi(t,r)=qt+\psi(r), \qquad X=\frac{q^2}{2h}-\frac{f\psi'^2}{2},
\end{eqnarray}
with $q$ being a constant.
Here a prime denotes the derivative with respect to the radial coordinate.

For the choice of the arbitrary functions
\begin{equation}
G_2(X)=-2\Lambda+2\eta X,~~~G_4(X)=\zeta+\beta X,~~~G_3(X)=G_5(X)=0, \label{dc2Gmn}
\end{equation}
the action can be expressed in the form~\cite{Kobayashi:2014eva}
\begin{equation}
S = \int d^4x \sqrt{-g} \left[ \zeta R - \eta \left( \nabla_\mu \phi \right)^2 + \beta G^{\mu\nu} \nabla_\mu \phi \nabla_\nu \phi - 2\Lambda \right], \label{ac}
\end{equation}
which has nonminimal derivative coupling to the Einstein tensor.
Here, we take $\zeta>0$ in accordance with GR and we assume $\beta\ne 0$.
For $\eta\neq 0$, the model~\eqref{ac} admits solutions in which the $\Lambda$-term is totally screened. The metric then is not asymptotically flat but rather dS with the effective cosmological constant proportional to $\eta/\beta$, since the scalar kinetic term becomes constant around the present time~\cite{Gubitosi:2011sg}. It offers an exciting opportunity to circumvent the cosmological constant problem.

As shown in Ref.~\cite{Babichev:2013cya}, BH solutions in the theory~\eqref{ac} possess very similar properties, governed by the following equations:
\begin{eqnarray}
&& f(r)  =  \frac{( \beta + \eta r^2) h}{\beta \left( rh\right)'},\label{f0} \\
&& h(r)  =  -\frac{\mu}{r} + \frac{1}{r} \int\frac{k(r)}{\beta + \eta r^2}dr, \label{h0}\\
&&\psi'^2=\frac{r}{( \beta + \eta r^2)^2 h^2}\bra{ q^2\beta (\beta+\eta r^2)h'- \frac{\zeta\eta+\Lambda\beta}{2}(r^2h^2)' }, \label{psi}
\end{eqnarray}
where $\mu$ plays the role of the BH mass and $k(r)$ is obtained from the following algebraic equation:
\begin{equation}
q^2\beta\left( \beta + \eta r^2\right)^2 -\brb{2\zeta\beta + \left( \zeta\eta-\Lambda\beta\right) r^2}k + C k^{3/2} =0. \label{k}
\end{equation}
Here, $C$ is an integration constant.
With these equations, the scalar kinetic term~$X$ is written as
\begin{equation}
X=\frac{(\zeta\eta+\Lambda\beta)r^2}{2\beta(\eta r^2+\beta)}+\fr{q^2(\eta r^2+\beta)}{2k(r)}. \label{generalX}
\end{equation}
If we assume $q\ne 0$, the system of equations~\eqref{f0}-\eqref{k} is obtained by combining $tt$-, $rr$- and $tr$-components of the field equations, while the scalar field equation is redundant.
On the other hand, if we set $q=0$, the $tr$-component becomes trivial and one has to use the scalar field equation instead\footnote{If $q=0$, there is another branch of solution, which may not be obtained analytically~\cite{Rinaldi:2012vy,Minamitsuji:2013ura}.}.
Therefore, below we consider the solutions for this system of equations both for $q\ne 0$ and $q=0$. 

If we take $q=0$ and assume $C\ne 0$, Eq.~\eqref{k} gives a nontrivial solution
\begin{equation}
k(r)=\fr{1}{C^2}\brb{ 2\zeta\beta + \left( \zeta\eta-\Lambda\beta\right) r^2}^2.
\end{equation}
Then, the metric functions~$f,h$ and the radial part of the scalar field~$\psi$ are successively obtained from Eqs.~\eqref{f0}-\eqref{psi}.
Depending on the signs of the model parameters~$\eta$ and $\beta$, the $q=0$ solutions are classified into four groups as follows.

\begin{enumerate}

\item {\bf The case $\eta\beta>0$}\\
We firstly consider the case $\eta\beta>0$. The metric function~$h$ can be found from Eqs.~\eqref{h0}, \eqref{k} as
\begin{equation}
h(r)=\frac{\beta(3\zeta\eta+\Lambda\beta)(\zeta\eta-\Lambda\beta)}{C^2\eta^2}-\frac{\mu}{r}+\frac{(\zeta\eta-\Lambda\beta)^2}{3C^2\eta}r^2+\frac{\beta(\zeta\eta+\Lambda\beta)^2 }{C^2\eta^2 }\frac{\arctan y}{y}, \label{met1}
\end{equation}
with $y\equiv r\sqrt{\eta/\beta}$.
The particular solutions of this kind (for specific choices of $C$) were obtained in Refs.~\cite{Rinaldi:2012vy,Babichev:2013cya,Anabalon:2013oea,Minamitsuji:2013ura}.
Note that Eq.~\eqref{met1} solves the background field equations even for $C^2<0$, i.e., for pure imaginary value of $C$. Such a possibility was considered in Ref.~\cite{Charmousis:2015aya}.
We see that the solution is asymptotically dS for 
\begin{equation}
C^2\eta<0 \label{ds}
\end{equation}
and anti-de Sitter (AdS) otherwise.

\item {\bf The case $\eta\beta<0$}\\
Now let us switch to the case $\eta\beta<0$.
The metric function~$h$ then reads
\begin{equation}
h(r)=\frac{\beta(3\zeta\eta+\Lambda\beta)(\zeta\eta-\Lambda\beta)}{C^2\eta^2}-\frac{\mu}{r}+\frac{(\zeta\eta-\Lambda\beta)^2}{3C^2\eta}r^2+\frac{\beta(\zeta\eta+\Lambda\beta)^2 }{C^2\eta^2}\cdot \fr{1}{2z}\ln \left|\fr{1+z}{1-z}\right|, \label{met2}
\end{equation}
with $z\equiv r\sqrt{|\eta/\beta|}$.
The solution is singular at $z=1$, i.e. $r=\sqrt{|\beta/\eta|}$, since the logarithmic term would diverge there, but this feature could be hidden behind the de Sitter horizon of the solution by adjusting $\beta/\eta$\footnote{The surface~$r=\sqrt{|\beta/\eta|}$ is not a coordinate singularity but a true singularity in the sense that the Ricci scalar~$R$ diverges there.}.
This can be achieved due to the interplay of the decaying dS term and the growing logarithm, which could give rise to a root (dS horizon) before $r=\sqrt{|\beta/\eta|}$.
We assume that the physically relevant region is included within $r<\sqrt{|\beta/\eta|}$ so that the last term in Eq.~\eqref{met2} is written in terms of the inverse hyperbolic tangent:
\begin{equation}
h(r)=\frac{\beta(3\zeta\eta+\Lambda\beta)(\zeta\eta-\Lambda\beta)}{C^2\eta^2}-\frac{\mu}{r}+\frac{(\zeta\eta-\Lambda\beta)^2}{3C^2\eta}r^2+\frac{\beta(\zeta\eta+\Lambda\beta)^2 }{C^2\eta^2}\fr{{\rm artanh}~z}{z}.
\end{equation}
Particular solutions of this kind were obtained in~\cite{Rinaldi:2012vy,Minamitsuji:2013ura}.

\item {\bf The case $\zeta\eta+\Lambda\beta=0$}\\
This is the degenerate case of (i) or (ii).
The metric functions~$h$ and $f$ simply read
\begin{equation}
f(r)=1-\fr{\mu}{r}+\fr{\eta}{3\beta}r^2,~~~h(r)=\fr{4\zeta^2\beta}{C^2}f(r).
\end{equation}
Therefore, redefining the time coordinate, one obtains the Schwarzschild-(A)dS solution.
On the other hand, the scalar field becomes trivial in this case.

\item {\bf The case $\eta=0$}\\
Finally, we discuss the case~$\eta=0$.
In this case, the action~\eqref{ac} falls into the so-called Fab Four theory, whose action is given by~\cite{Charmousis:2011bf}
\begin{equation}
S=\int d^4x\sqrt{-g}\bra{\mL_{\rm John}+\mL_{\rm Paul}+\mL_{\rm George}+\mL_{\rm Ringo}-2\Lambda}, \label{fab4}
\end{equation}
where
\begin{eqnarray}
\mL_{\rm John}&=&V_{\rm John}(\phi)G^{\mu\nu}\na_\mu\phi\na_\nu\phi, \\
\mL_{\rm Paul}&=&V_{\rm Paul}(\phi)P^{\mu\nu\lambda\sigma}\na_\mu\phi\na_\lambda\phi\na_\nu\na_\sigma\phi, \\
\mL_{\rm George}&=&V_{\rm George}(\phi)R, \\
\mL_{\rm Ringo}&=&V_{\rm Ringo}(\phi)\bra{R^2-4R_{\mu\nu}R^{\mu\nu}+R_{\mu\nu\lambda\sigma}R^{\mu\nu\lambda\sigma}}.
\end{eqnarray}
Here, $P^{\mu\nu\lambda\sigma}$ is the double dual of the Riemann tensor.
The action~\eqref{ac} with $\eta=0$ amounts to the choice
\begin{equation}
V_{\rm John}=\beta,~~~V_{\rm Paul}=0,~~~V_{\rm George}=\zeta,~~~V_{\rm Ringo}=0.
\end{equation}
The Fab Four theory represents the unique subset of Horndeski theories that allows for the existence of a consistent self-tuning mechanism on Friedmann-Lema\^itre-Robertson-Walker (FLRW) background, yielding the Minkowski metric as a resultant spacetime\footnote{For such models without assuming a Minkowski vacuum, see Refs.~\cite{Martin-Moruno:2015bda,Starobinsky:2016kua}.}.
This self-tuning requires that the vacuum cosmological constant~$\Lambda$ should not impact the curvature.
Thus, whatever the value of $\Lambda$ is, we can have the Minkowski spacetime as a solution of the theory, while the theory admits nontrivial cosmology as well. The idea is that the cosmological field equations should be dynamical, with the Minkowski solution corresponding to some sort of fixed point. In other words, once we are on the Minkowski solution, we stay there, otherwise we evolve to it dynamically.

In the present case of $\eta=0$, we get the following expression for the metric function~$h$:
\begin{equation}
h(r)=\frac{4\zeta^2\beta}{C^2}-\frac{\mu}{r}-\frac{4\zeta\Lambda\beta}{3C^2}r^2 + \frac{\Lambda^2\beta}{5C^2}r^4, \label{bhj}
\end{equation}
which amounts to the solution given in Ref.~\cite{Anabalon:2013oea}.

\end{enumerate}

\section{Stability of black hole solutions} \label{bhs}

\subsection{Solutions with linearly time-dependent scalar hair}

Here, we consider a generic shift- and reflection-symmetric Horndeski theory~\eqref{lag} and discuss the stability of solutions with nonzero scalar velocity charge:~$q\ne 0$.
In Refs.~\cite{Ogawa:2015pea,Takahashi:2016dnv}, the authors calculated the quadratic action that governs the dynamics of odd-parity perturbations.
From the requirement that the kinetic/gradient energy should be positive, they obtained the necessary conditions for stability as
\begin{eqnarray}
&&\mathcal{F}>0, \qquad \mathcal{G}>0, \qquad \mathcal{H}>0, \label{stab} \\ 
&&\mathcal{F}\equiv 2\left( G_4-\frac{q^2}{h}G_{4X}\right), \label{stab1} \\
&&\mathcal{G}\equiv 2\left[G_4+\left(\frac{q^2}{h}-2X\right)G_{4X}\right], \label{stab2} \\
&&\mathcal{H}\equiv 2(G_4-2XG_{4X}).
\end{eqnarray}
Note that the conditions above apply for $q=0$ solutions as well.
To show the instability, we investigate the behavior of the variables~$\mathcal{F},\mathcal{G}$ near the horizon~($h\simeq 0$). Since $X$ is finite at the horizon for physically relevant solutions, the terms with $q^2/h$ dominate in Eqs.~\eqref{stab1} and \eqref{stab2}. Therefore, we obtain~\cite{Ogawa:2015pea,Takahashi:2016dnv}
\begin{equation}
\mathcal{F}\mathcal{G}\approx -\left(\frac{2q^2}{h}G_{4X}\right)^2<0, \label{FG}
\end{equation}
if $G_{4X}$ takes some finite value near the horizon. 
Equation~\eqref{FG} means that the two of the stability conditions~$\mathcal{F}>0$ and $\mathcal{G}>0$ cannot be satisfied simultaneously for $q\ne 0$, leading to ghost/gradient instability.
Particularly, for the theory~\eqref{ac}, $G_{4X}=\beta\ne 0$ and thus BHs with $q\ne 0$ always suffer from the instability.

However, this is not the case if $G_{4X}$ is vanishing.
As was proposed in Ref.~\cite{Takahashi:2016dnv}, this can be obviously realized by setting $G_4$ to be constant, which amounts to GR plus a noncanonical scalar field.
Furthermore, there is another way to circumvent the instability mentioned above:
having $G_{4X}=0$ at least in the vicinity of the horizon for a specific background solution.
While the authors of Ref.~\cite{Takahashi:2016dnv} gave such a solution with $X=0$, here we present  a new example with $X$ being a nonzero constant.
Consider the following model:
\begin{equation}
G_2(X)=-2\Lambda+2\eta X,~~~G_4(X)=\zeta+\beta X+\fr{\gamma}{2}X^2,~~~G_3(X)=G_5(X)=0,
\end{equation}
where the difference from Eq.~\eqref{dc2Gmn} is the quadratic term in $G_4(X)$ with $\gamma\ne 0$.
If we choose $\Lambda$ as
\begin{equation}
\Lambda=\fr{\eta}{4\beta\gamma}(2\zeta\gamma-5\beta^2),
\end{equation}
the following configuration solves the the background field equations listed in Ref.~\cite{Ogawa:2015pea}:
\begin{equation}
f(r)=1-\fr{\mu}{r}-\fr{\eta}{6\beta}r^2,~~~h(r)=-\fr{\gamma q^2}{2\beta}f(r),~~~\psi'^2=-\fr{2\beta}{\gamma}\fr{1-f}{f^2}, \label{stableXsquared}
\end{equation}
which is the Schwarzschild-(A)dS solution up to redefinition of the time coordinate.
This configuration satisfies
\begin{eqnarray}
G_{4X}&=&\beta+\gamma X=0, \label{Xconst}
\end{eqnarray}
and thus
\begin{equation}
\mF=\mG=\mH=\zeta-\fr{\beta^2}{2\gamma}.
\end{equation}
Note that Eq.~\eqref{Xconst}, i.e., $X=-\beta/\gamma={\rm const.}$ holds for any $r$.
Therefore, for $\zeta>\beta^2/(2\gamma)$, the solution \eqref{stableXsquared} evades the aforementioned instability despite the nontrivial $X$-dependence of $G_4$.

\subsection{Solutions with static scalar hair}

We now switch to $q=0$ solutions. For concreteness, below we restrict ourselves to the particular theory~\eqref{ac}.
The scalar kinetic term~\eqref{generalX} then reads
\begin{eqnarray}
X= \frac{(\zeta\eta+\Lambda\beta)r^2}{2\beta(\eta r^2+\beta)}. \label{x}
\end{eqnarray}
For the model~\eqref{ac}, the conditions~\eqref{stab} read \cite{Kobayashi:2014wsa}
\begin{eqnarray}
&&\frac{ r^2  (\zeta\eta+\Lambda \beta)  }{2(\eta r^2  + \beta)}  + \zeta >0, \label{stab1_1}\\
-&&\frac{ r^2  (\zeta\eta+\Lambda \beta)  }{2(\eta r^2  + \beta)}  + \zeta >0. \label{stab2_1}
\end{eqnarray}

Next, we consider the stability conditions for even-parity perturbations.
Contrary to odd-parity perturbations, now there are two propagating degrees of freedom: gravitational and scalar waves.
Computing the quadratic action for the even-parity perturbations, the authors of Ref.~\cite{Kobayashi:2014wsa} obtained the stability conditions for $q=0$ solutions\footnote{The stability conditions for even-parity perturbations in $q\ne 0$ case have not been known.}.
For the model in question, these conditions read
\begin{align}
\fr{f(2r\mH+\Xi\psi')}{hr^2\mH^2}\brb{\fr{hr^4\mH^4}{f(2r\mH+\Xi\psi')^2}}'-\mF&>0, \label{evensta1} \\
2r^2\Ga\mH\Xi\psi'^2-\mG\Xi^2\psi'^2-\fr{4r^4}{f}\Sigma\mH^2&>0, \label{evensta2}
\end{align}
where
\begin{align}
\Xi&\equiv 4rf\psi'(G_{4X}+2XG_{4XX}), \\
\Ga&\equiv \bra{\fr{4}{r}+\fr{2h'}{h}}f\psi'(G_{4X}+2XG_{4XX}), \\
\Sigma&\equiv XG_{2X}+2X^2G_{2XX}+2\bra{\fr{1-f}{r^2}-\fr{f}{r}\fr{h'}{h}}XG_{4X}\\
&\quad +4\bra{\fr{1-4f}{r^2}-\fr{4f}{r}\fr{h'}{h}}X^2G_{4XX}-\fr{8f}{r}\bra{\fr{1}{r}+\fr{h'}{h}}X^3G_{4XXX}.
\end{align}
These conditions provide the positive squared propagation speed of the scalar wave and the no-ghost condition.
For the solutions of the master equations~\eqref{f0}-\eqref{k}, the conditions~\eqref{evensta1}, \eqref{evensta2} respectively read
\begin{align}
-\fr{4r^4(\zeta\eta+\Lambda\beta)^2\brb{\eta(\zeta\eta+3\Lambda\beta)r^4-2\beta(2\zeta\eta-3\Lambda\beta)r^2-8\zeta\beta^2}}{(\eta r^2+\beta)^2\brb{(\zeta\eta+3\Lambda\beta)r^2-2\zeta\beta}^2}&>0, \label{evensta1_1} \\
\fr{16r^6(\zeta\eta+\Lambda\beta)^2\brb{(\zeta\eta-\Lambda\beta)r^2+2\zeta\beta}^3}{C^2h(\eta r^2+\beta)^4}&>0. \label{evensta2_1}
\end{align}
Assuming $\zeta\eta+\Lambda\beta\ne 0$, these simplify as
\begin{align}
\eta(\zeta\eta+3\Lambda\beta)r^4-2\beta(2\zeta\eta-3\Lambda\beta)r^2-8\zeta\beta^2&<0, \label{evensta1_2} \\
\fr{(\zeta\eta-\Lambda\beta)r^2+2\zeta\beta}{C^2}&>0. \label{evensta2_2}
\end{align}

Now that we have clarified the stability conditions for both odd- and even-parity perturbations, we are in position to discuss the stability of the BH solutions presented in Sec.~\ref{background}.

\begin{enumerate}
\item {\bf The case $\eta\beta>0$}\\
We now sum up all the stability conditions. Firstly, for odd-parity perturbations, the conditions~\eqref{stab1_1} and \eqref{stab2_1} yield
\begin{eqnarray}
\beta(3\zeta\eta+\Lambda\beta)r^2+2\zeta\beta^2>0,&& \\
\beta(\zeta\eta-\Lambda\beta)r^2+2\zeta\beta^2>0.&&
\end{eqnarray}
If we require these inequalities are satisfied for any $r\ge 0$, we have the following constraints on the model parameters:
\begin{equation}
\beta(3\zeta\eta+\Lambda\beta)\ge 0,~~~\beta(\zeta\eta-\Lambda\beta)\ge 0.
\end{equation}
Next, we require even-parity perturbations to be stable.
The condition~\eqref{evensta1_2} implies\footnote{The necessary and sufficient condition for a quartic inequality~$ar^4+br^2+c>0~(a\ne 0, b^2\ge 4ac)$ to be satisfied for any real $r$ is $a\ge 0$, $b\ge 0$, $c>0$.}
\begin{equation}
\eta(\zeta\eta+3\Lambda\beta)\le 0,~~~\beta(2\zeta\eta-3\Lambda\beta)\ge 0,
\end{equation}
while the condition~\eqref{evensta2_2} requires
\begin{equation}
\fr{\zeta\eta-\Lambda\beta}{C^2}\ge 0,~~~\fr{\beta}{C^2}>0.
\end{equation}

For $\eta>0$ and $\beta>0$, the entire set of stability conditions is satisfied if
\begin{equation}
3\zeta\eta+\Lambda\beta\ge 0,~~~\zeta\eta+3\Lambda\beta\le 0,~~~C^2>0. \label{stab(i)_1}
\end{equation}
For $\eta<0$ and $\beta<0$, the same stability conditions become
\begin{equation}
3\zeta\eta+\Lambda\beta\le 0,~~~\zeta\eta+3\Lambda\beta\ge 0,~~~C^2<0. \label{stab(i)_2}
\end{equation}
As noted below Eq.~\eqref{ds}, the solution~\eqref{met1} with such parameters are asymptotically AdS.
Such a solution may be interesting in the context of AdS/CFT correspondence and brane cosmology, along the lines of Ref.~\cite{Birmingham:2001dq}.

\item {\bf The case $\eta\beta<0$}\\
Since the physically relevant region is included within $r<\sqrt{|\beta/\eta|}$ in this case, it suffices if the stability conditions are satisfied in this restricted region.
However, the stability conditions~\eqref{stab1_1}, \eqref{stab2_1} for odd-parity perturbations cannot be satisfied simultaneously.
This is because the left-hand sides of these inequalities have different sign near $r=\sqrt{|\beta/\eta|}$ unless $\zeta\eta+\Lambda\beta\ne0$: One of them goes to positive infinity, while the other goes to negative infinity.
Thus, this type of solutions is always plagued by ghost/gradient instability even at the level of linear odd-parity perturbations.

\item {\bf The case $\zeta\eta+\Lambda\beta=0$}\\
In this case, the scalar profile becomes constant and the apparent metric is just Schwarzschild-(A)dS.
It is notable that the left-hand sides of Eqs.~\eqref{evensta1_1}, \eqref{evensta2_1} are vanishing, which may originate from the strong coupling of the perturbation corresponding to the scalar wave on the background solution~\eqref{bhj}.
For the other modes, i.e., gravitational wave, there is no signal of instability.

\item {\bf The case $\eta=0$}\\
For $\eta=0$, the stability conditions read
\begin{equation}
\Lambda r^2+2\zeta >0,~~~-\Lambda r^2+2\zeta >0,~~~3\Lambda r^2-4\zeta<0,~~~\fr{\beta}{C^2}>0.
\end{equation}
The first three conditions will be violated at a large enough $r$ unless $\Lambda= 0$.
Conversely, if one takes $\Lambda=0$ and $\beta/C^2>0$, where the solution~\eqref{bhj} becomes of Schwarzschild form, then all the above conditions are satisfied.
Thus, the BH~\eqref{bhj} is unstable or represents a trivial configuration.
This indicates that the John term of the Fab Four action cannot be used in isolation, which is also true for the other terms: 
The Paul term has problems with describing neutron stars~\cite{Maselli:2016gxk}, while  with $V_{\rm George}(\phi)$ and $V_{\rm Ringo}(\phi)$ being constant the model falls into GR.

\end{enumerate}

The result are summarized in Table~\ref{tab}.
\begin{table}
	\begin{tabular}{|l||c|c|}
	\hline
	\multicolumn{1}{|c||}{Case} & \multicolumn{1}{|c|}{~~Nontrivial solution~~} & \multicolumn{1}{c|}{Stability conditions} \\ \hline\hline
	~~(i) $\eta\beta>0$ & yes & \multicolumn{1}{r|}{~~$\eta>0$,~~$3\zeta\eta+\Lambda\beta\ge 0$,~~$\zeta\eta+3\Lambda\beta\le 0$,~~$C^2>0$~~} \\
	 &  & \multicolumn{1}{r|}{~~or~~$\eta<0$,~~$3\zeta\eta+\Lambda\beta\le 0$,~~$\zeta\eta+3\Lambda\beta\ge 0$,~~$C^2<0$~~} \\ \hline
	~~(ii) $\eta\beta<0$ & yes & never stable \\ \hline
	~~(iii) $\zeta\eta+\Lambda\beta=0$~~ & no & always stable \\ \hline
	~~(iv) $\eta=0$ & no & $\Lambda=0$,~~$\beta/C^2>0$ \\
	\hline
	\end{tabular}
	\caption{Summary of the BH solutions for specific parameter regions. The second column shows whether each case has a nontrivial solution or not. The stability conditions are listed in the third column.}
	\label{tab}
\end{table}
To sum up, while the BH solutions for the case $\eta\beta<0$ with $\zeta\eta+\Lambda\beta\ne 0$ are unstable, there is some parameter region that could evade instability for the other cases.
Our result for the case $\eta\beta>0$ is consonant with that in Ref.~\cite{Kobayashi:2014wsa}, where the authors considered only large-$r$ behavior of stability conditions to obtain the parameter region.
Note that the boundary of physically relevant parameter region could be extended even further:
there is still possibility that the instability region is hidden behind the horizon.

\section{Conclusions} \label{conc}

Although BH solutions with linearly time-dependent scalar hair in shift- and reflection-symmetric Horndeski theories generally suffer from ghost/gradient instability, there are still some solutions that circumvent the instability.
Apart from the trivial loophole with $G_4$ being constant, we proposed a new solution possessing a linearly time-dependent scalar profile, with the model parameters fine-tuned to avoid the instability.

For solutions with a static scalar profile where the generic instability is absent, we focused on a theory with nonminimal derivative coupling to the Einstein tensor and refined the stability conditions obtained in Ref.~\cite{Kobayashi:2014wsa}.
The resultant parameter region is summarized in Table~\ref{tab}, which provides the BH stability for any $r$.
It should be noted that the stability conditions we discussed are only necessary conditions for the mode stability.
To complete the mode stability analysis, one must investigate the nature of the potential part in the quadratic action as in Ref.~\cite{Takahashi:2016dnv}.
We leave these issues for the future work.

\section{Acknowledgements}
D.A.T. is supported by Russian Foundation for Basic Research via grant RFBR \No 16-02-00682, by Act 211 of the Russian Federation Government, agreement \No 02.A03.21.0006, by the Ministry of Education and Science project \No 5719. 
K.T. would like to thank Teruaki Suyama for useful discussions.

\bibliography{mybib}
\end{document}